\documentclass[letterpaper,journal]{IEEEtran}
\usepackage{amsmath,amsfonts}
\usepackage{algorithmic}
\usepackage{algorithm}
\usepackage{array}
\usepackage[caption=false,font=normalsize,labelfont=sf,textfont=sf]{subfig}
\usepackage{textcomp}
\usepackage{stfloats}
\usepackage{url}
\usepackage{verbatim}
\usepackage{graphicx}
\usepackage{cite}
\usepackage{booktabs}
\usepackage{amssymb}
\usepackage{xcolor}
\usepackage{comment}
\usepackage{orcidlink}
\hypersetup{hidelinks}
\usepackage{multirow}
\usepackage{balance}

\begin{document}

\title{REACH: Controller-Managed Long-Span ECC for HBM AI Inference}

\author{Rui Xie\orcidlink{0000-0003-3177-5071}, Yunhua Fang\orcidlink{0009-0009-4718-8825}, Asad Ul Haq\orcidlink{0009-0003-7975-0102}, Linsen Ma\orcidlink{0009-0000-8535-7911}, Sanchari Sen\orcidlink{0000-0003-0080-2882}, Swagath Venkataramani\orcidlink{0000-0002-0470-6364}, Liu Liu\orcidlink{0000-0003-0792-8146}, Tong Zhang\orcidlink{0009-0009-8005-0043}

\thanks{Rui Xie, Yunhua Fang, Asad Ul Haq, Linsen Ma, Liu Liu, and Tong Zhang are with Rensselaer Polytechnic Institute, Troy, NY 12180 USA.}

\thanks{Sanchari Sen and Swagath Venkataramani are with IBM T.J. Watson Research Center, Yorktown Heights, NY 10598 USA.} 

}

\maketitle


\begin{abstract}
High-Bandwidth Memory (HBM) cost motivates stronger controller protection that can support a wider range of device error rates. Long-span error-correcting codes provide stronger protection at a comparable code rate, but a direct implementation couples small accesses to span-wide state and requires costly decoding at HBM bandwidth. Read-dominated LLM decode offers a favorable setting: sequential reads support span aggregation, while sparse writes limit parity-update traffic. This paper presents REACH, a controller microarchitecture that uses established inner codes to correct common errors and identify unresolved chunks, reserving a long outer code for known-erasure repair. Differential parity bounds write traffic, and a co-designed endpoint preserves 32\,B transactions without an extra data burst. Ramulator2 sustains 1.88\,TB/s of application traffic at the highest error stress, while separate full-interface sizing supports a 2.69\,TB/s application target using ASAP7-synthesized kernels. At this analytical target, REACH's nominal composition uses 55.8\% less controller area and 57.7\% less modeled power than the evaluated mean-work direct-long design, showing the benefit of reserving long-span recovery for exceptional requests.
\end{abstract}


\section{Introduction}
High-Bandwidth Memory (HBM) has become indispensable to modern AI inference because its multi-terabyte-per-second bandwidth sustains model-weight and key-value-state movement within a practical energy budget. HBM also remains 5--10$\times$ more expensive than conventional dynamic random-access memory (DRAM) in \$/GB~\cite{Koch2024TheMW}, making capacity a first-order system cost. The emergence of NVHBM makes HBM base dies and interfaces explicit co-design points for custom AI accelerators, creating a timely opportunity to revisit how reliability work is divided between the endpoint and controller~\cite{nvidia-nvhbm-2026}. NAND storage provides a precedent for using stronger controller coding and selective recovery as device reliability changes~\cite{zhao2013ldpc}. This motivates stronger HBM controller protection that preserves accelerator-visible service across a wider range of device error rates.

HBM requests are fine grained. Transfer-local protection fits this interface because each request can be resolved independently, but its correction strength is limited by the redundancy carried with that request. Adding more local redundancy consumes capacity and transport resources on every access. A longer codeword can pool the same protection budget across many requests and provide substantially stronger correction.

Autoregressive LLM decode creates favorable conditions for using this larger span. It repeatedly streams weights and reads a large key-value cache while appending much less new state. Its traffic is therefore read dominated, highly sequential, and governed by sustained memory service. Contiguous requests can share span work, while detected exceptions can be handled away from ordinary demand service without relaxing correctness.

These properties make long-span ECC a natural strong-protection baseline. Ideal coalescing aggregates sequential requests into protected spans and reduces traffic. Sparse accesses still touch span-wide parity state. The controller must check every admitted span and provision long-code recovery for the fraction that fails this check.
\begin{figure}[htbp]
    \centering
    \includegraphics[width=\linewidth]{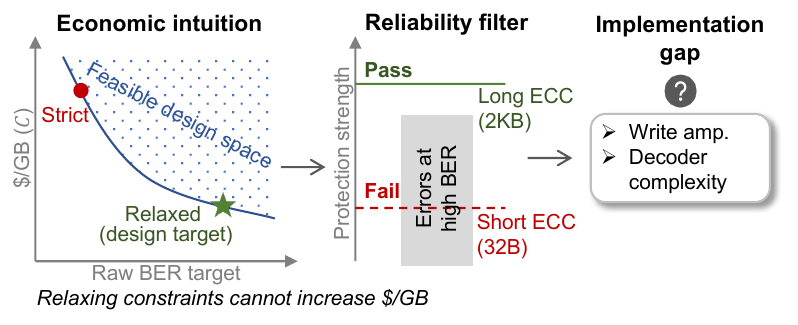}
    \caption{\textbf{Why controller-managed long-span ECC for AI inference.}
    \textbf{Left:} HBM cost pressure motivates exploring less restrictive device-reliability targets.
    \textbf{Middle:} Long-span protection supplies the correction strength required as device errors increase.
    \textbf{Right:} A direct implementation must handle sparse-write span traffic and provision long-code checking and error-dependent recovery at HBM bandwidth.}
    \label{fig:problem_overview}
\end{figure}

Fig.~\ref{fig:problem_overview} connects the device opportunity to the controller problem. The left region motivates a broader device design space. The middle region shows why that opportunity requires stronger protection. The right region identifies the traffic and hardware costs of applying a long code directly at the controller. Sequential requests can share complete-span traffic, while sparse writes still expose the 64-to-1 granularity difference and every protected span still requires long-code processing.

Even with ideal 2\,KB coalescing, a 3.35\,TB/s data stream presents up to $1.64\times10^9$ protected spans/s before outer-parity traffic. Each span must be checked, and corrupted spans must pass through the complete recovery pipeline. Faster arithmetic reduces the cost of individual operations, but HBM bandwidth still demands extensive replication of that pipeline. Section~\ref{sec:opportunity} quantifies the decoder work that creates this pressure.

The controller can change this scaling by changing the information delivered to long-span recovery. If an earlier 32\,B stage reports that a request is unresolved, the request address already identifies the affected region of the outer codeword. The long code can then recover a declared erasure after rejection, while ordinary requests complete without entering the outer path. This organization reduces both the work required by each outer invocation and the rate at which outer hardware is invoked.

This leads naturally to a systems question: can a controller preserve 32\,B common-case service while reserving a long code for detected exceptions? \textbf{REACH} (Reliability Extension Architecture for Controller-managed HBM) uses established inner and outer coding to correct common errors locally and explicitly reject unresolved chunks. A rejected chunk becomes an address-derived erasure in a single monolithic outer codeword, which is invoked only on the exceptional path. Differential parity bounds write traffic, while accepted and corrected requests retain 32\,B completion.

We evaluate this workflow with a public-model-based traffic model, Ramulator2 command simulation, and ASAP7-synthesized kernels. Ramulator2 sustains 70\% HBM offered load (2.35\,TB/s total), including 1.88\,TB/s of application bandwidth at the highest traffic stress. At the separate analytical 2.69\,TB/s application target, REACH's nominal composition uses 55.8\% less controller area and 57.7\% less modeled power than the mean-work direct-long design. Full-radius direct-long provisioning is reported separately as a hardware and activity sensitivity.

This paper makes 3 contributions:

\noindent $\bullet$ \textbf{HBM feasibility diagnosis.} We quantify the span-rate checking and error-dependent recovery resources that direct long-code protection requires even with ideal span aggregation.

\noindent $\bullet$ \textbf{Controller workflow and microarchitecture.} We organize established coding primitives into a 32\,B inner decision, exceptional known-erasure repair, and differential-parity workflow with explicit state, ordering, completion, and endpoint contracts.

\noindent $\bullet$ \textbf{HBM-scale feasibility evidence.} Reliability analysis, model-derived LLM-decode traffic, finite-queue service, Ramulator2 command simulation, and synthesized critical kernels quantify the workflow at HBM service rates.
\section{Background, Opportunity, and Requirements}
\label{sec:background}

\subsection{HBM and LLM Decode Context}
\label{sec:hbm-constraints}

HBM combines a 32\,B accelerator-visible demand transaction with multi-TB/s aggregate bandwidth~\cite{JEDEC_JESD270_4_2025,JEDEC_JESD238B01_2025}. The memory controller resides on the accelerator and connects the L2 cache or network-on-chip to the HBM scheduler and physical interface. Any controller-managed protection must preserve the address, ordering, and completion semantics of this path.

LLM decode repeatedly streams model weights and key-value state while appending a much smaller amount of new state. Contiguous accesses within tensors and key-value blocks allow aligned requests to share span work. We therefore give the direct-long baseline ideal coalescing within these runs and separately evaluate random requests and sparse writes.

Public HBM specifications do not expose a common on-die parity matrix or protected word. We therefore model a vendor-neutral 32\,B on-die-ECC control that corrects at most 1 affected 16-bit group as a separate comparison. REACH mode bypasses conventional on-die correction and moves active ECC to the accelerator-side controller. Throughout the paper, $p_b$ denotes the modeled HBM output bit error rate (BER), the probability that a bit returned by the co-designed endpoint differs from its intended value before controller correction. The errors originate in the HBM device and are observed at the controller. Public sources do not provide a common mapping from cell-level errors through sensing, remapping, and interface transfer to this output BER.

\subsection{Long-Span Coding Opportunity}
\label{sec:longer-ecc-necessary}

The comparison uses shortened Reed--Solomon (RS) codes over Galois fields (GF).

\begin{figure}[htbp]
  \centering
  \includegraphics[width=\linewidth]{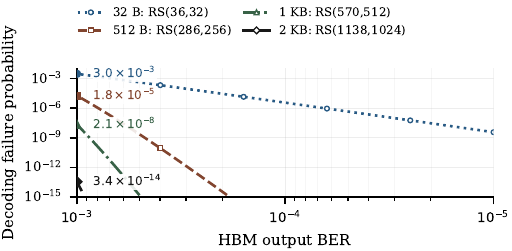}
  \caption{Analytical decoding failure probability for monolithic RS codewords from 32\,B to 2\,KB under the representative near-0.9 target-rate rule. The horizontal axis runs from the highest evaluated HBM output BER at left toward lower values. Filled markers report each value at $p_b=10^{-3}$. The 32\,B fixed-rate reference uses GF$(2^8)$, while larger spans use GF$(2^{16})$. The 2\,KB point is the evaluated outer RS(1138,1024).}
  \label{fig:rs_fail_vs_n}
\end{figure}

Fig.~\ref{fig:rs_fail_vs_n} shows the coding opportunity behind a longer protection span. Each code follows the same target-rate rule as closely as its symbol count allows. The 32\,B reference is RS(36,32) over GF$(2^8)$ and is separate from the evaluated REACH inner code. The 512\,B, 1\,KB, and 2\,KB points use RS(286,256), RS(570,512), and RS(1138,1024) over GF$(2^{16})$. At $p_b=10^{-3}$, their analytical failure probabilities are $3.0\times10^{-3}$, $1.8\times10^{-5}$, $2.1\times10^{-8}$, and $3.4\times10^{-14}$. Under the shared rate rule, increasing the span from 32\,B to 2\,KB reduces analytical failure by approximately 11 orders of magnitude.

\subsection{Why Direct Long ECC Does Not Fit}
\label{sec:opportunity}

The conventional direct-long baseline protects each aligned 2\,KB data span with the same monolithic outer code evaluated for REACH. It receives ideal coalescing within contiguous tensor and key-value runs. A sparse 32\,B update still changes parity shared by 64 native requests, so the controller must obtain old data and span parity beyond the new payload. Sequential aggregation reduces this traffic cost without changing the decoder that processes each span.

A conventional Reed--Solomon decoder begins by computing syndromes to determine whether a codeword is valid. A nonzero syndrome invokes key-equation solving, searches the codeword for unknown error positions, computes correction magnitudes, applies the corrections, and checks the residual syndrome. A longer codeword expands the search domain and the associated arithmetic. Serving HBM aggregate bandwidth then requires deeper pipelines, wider datapaths, or more replicated decoders.

\begin{figure}[htbp]
    \centering
    \includegraphics[width=.9\linewidth]{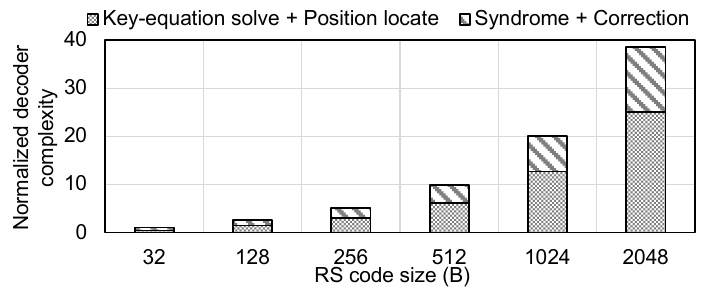}
    \caption{Normalized analytical operator demand of a conventional RS decoder. Syndromes check each codeword. Key-equation solving, position location, and correction handle nonzero syndromes. The plotted categories group syndrome with correction and key-equation solving with position location. Values are normalized to the 32\,B code.}
    \label{fig:codeword-complexity}
\end{figure}

Fig.~\ref{fig:codeword-complexity} groups the arithmetic into syndrome/correction and key-equation/position-location costs. These groups illustrate code-length scaling rather than invocation frequency: syndrome checking runs for every span, while recovery follows a nonzero syndrome. Table~\ref{tab:ppa-summary} applies these different invocation rates in a concrete comparison based on mapped kernels.

This analysis yields 4 design requirements: 32\,B common service, explicit locations for exceptional long-span recovery, parity updates that preserve unchanged data, and request ownership through completion or detected failure. Together, they connect coding strength to traffic, hardware demand, and controller correctness.

A stronger local code addresses only the 32\,B common-service requirement because its correction strength remains bounded by the metadata carried with each transfer. A faster direct-long decoder reduces latency or replication but still discovers error positions within every corrupted span. Satisfying all 4 requirements calls for a controller organization that corrects common errors locally, reports unresolved locations explicitly, and invokes long-span recovery only after rejection.


\section{REACH Controller Workflow and Microarchitecture}
\label{sec:method}

REACH organizes established coding primitives for HBM service. A short inner code classifies each returned 32\,B \emph{chunk} as accepted, corrected, or rejected. The request address maps a rejected chunk to its fixed position within a 64-chunk, 2\,KB protected \emph{span}. The controller can therefore present the unresolved chunk as a known outer-code erasure and invoke long-span recovery only after rejection.

For data chunk $j$, where $0\le j<64$, rejection identifies outer symbols $16j$ through $16j+15$. The same address rule locates every data record fetched for repair. Each full parity record maps to 16 parity-symbol positions, and the final parity beat contributes only 2 valid symbols. These mappings construct the erasure mask before outer decoding and remove long-code position search from the exceptional path.

\subsection{Placement and Interfaces}
\label{sec:microarch}

REACH is added to the distributed HBM memory-controller partitions already present on the accelerator die. It is not a separate ASIC. The upstream L2, network-on-chip (NoC), or memory fabric sends ordinary memory requests to these partitions. REACH schedules application, repair, and parity requests before the existing HBM command scheduler and physical interface (PHY). Returned records cross the channel interface as 32\,B of data and 6\,B of inner metadata.

\begin{figure}[htbp]
    \centering
    \includegraphics[width=\linewidth]{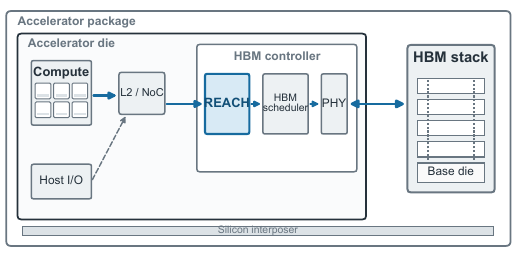}
    \caption{System placement of REACH. REACH extends each accelerator-die HBM controller partition between the upstream L2/NoC and the existing HBM scheduler and PHY. It reaches the co-packaged HBM stacks through the native in-package memory path. Host I/O remains outside this path.}
    \label{fig:system-placement}
\end{figure}

Fig.~\ref{fig:system-placement} places REACH within the existing accelerator memory path. Compute requests pass through the L2/NoC to the HBM controller, where REACH classifies and schedules application, repair, and parity traffic before the existing scheduler and PHY. Each controller partition reaches a co-packaged HBM stack through the native package interconnect, while through-silicon vias connect the stack's base die and DRAM dies. External host interfaces terminate at the accelerator I/O complex and do not carry ordinary accelerator-to-HBM requests. REACH therefore uses the same wide in-package path that provides the 3.35\,TB/s aggregate bandwidth.

\begin{table}[htbp]
    \centering
    \caption{Co-designed endpoint contract.}
    \label{tab:sideband-feasibility}
    \scriptsize
    \setlength{\tabcolsep}{2pt}
    \renewcommand{\arraystretch}{1.0}
    \begin{tabular}{@{}p{0.25\linewidth}p{0.68\linewidth}@{}}
        \toprule
        \textbf{Requirement} & \textbf{Contract} \\
        \midrule
        Accelerator interface & 32\,B demand transaction with unchanged address, ordering, and completion \\
        BL8 record & 4 byte-wise bus-inversion indicators and 2 ECC/severity I/Os carry 6 metadata bits beside 32 data bits in each unit interval \\
        Endpoint storage & Co-indexed data and metadata arrays selected by the same row and column command \\
        Mode lifecycle & Quiescent boot-time entry, standard inversion and on-die correction disabled, retraining, and no mixed runtime semantics \\
        \bottomrule
    \end{tabular}
\end{table}

The mode is selected at boot while the channel is quiescent. HBM3 associates 4 byte-wise bus-inversion indicators and 2 ECC/severity I/Os with each 32-bit pseudo-channel~\cite{JEDEC_JESD238B01_2025}. REACH disables those functions and, in BL8 unit interval $t$, uses the 6 bidirectional signals for metadata bits $6t$ through $6t+5$. Eight intervals transfer all 48 metadata bits beside the 256 data bits with the original command, burst, and strobes. Co-indexed arrays on the DRAM dies store data and metadata, while the base die aggregates their atomic 32\,B plus 6\,B record. This widens internal record movement by 18.75\% and requires endpoint storage, datapath, mode, and training changes, but adds no external command, DQ burst, or package pin. At the evaluated code rates, 71.11\,GB of protected demand capacity occupies 95\,GB of physical endpoint storage. Section~\ref{sec:e2e-ecc} gives the corresponding break-even condition. The design does not assume an unmodified commodity HBM device.

\subsection{Coding and Storage Organization}

Each chunk uses a systematic inner RS(38,32) code over GF$(2^8)$ with 32 data bytes and 6 parity bytes. The outer tier uses 1 shortened systematic RS(1138,1024) codeword over GF$(2^{16})$. Its 1024 2-byte data symbols represent the 64 data chunks, and 114 parity symbols provide the outer protection. The 228\,B of logical parity occupies 8 physical 32\,B beats, with only 2 valid symbols in the final beat. A rejected data chunk declares 16 erasures, so the outer code can recover up to 7 full-chunk-equivalent erasures. All data and parity symbols remain in the same monolithic codeword.

\noindent\textbf{Design-point rationale.}\quad
The 6\,B endpoint budget fixes the evaluated inner code. We use a radius-2 decoder to correct common outcomes and explicitly reject unresolved chunks. The 2\,KB outer span covers 64 native requests, tolerates 7 full-chunk-equivalent erasures, and bounds the amount of repair traffic. These choices define the representative design point used throughout the paper.

The systematic layout preserves the address order of all 64 data chunks. A request address directly yields the span identifier and chunk index, while the span identifier selects the associated physical parity record. Padding in the final parity beat does not enter the codeword. The controller therefore derives every data and parity erasure position from the request address or repair-read index without an additional metadata lookup.

\begin{figure}[htbp]
    \centering
    \includegraphics[width=\linewidth]{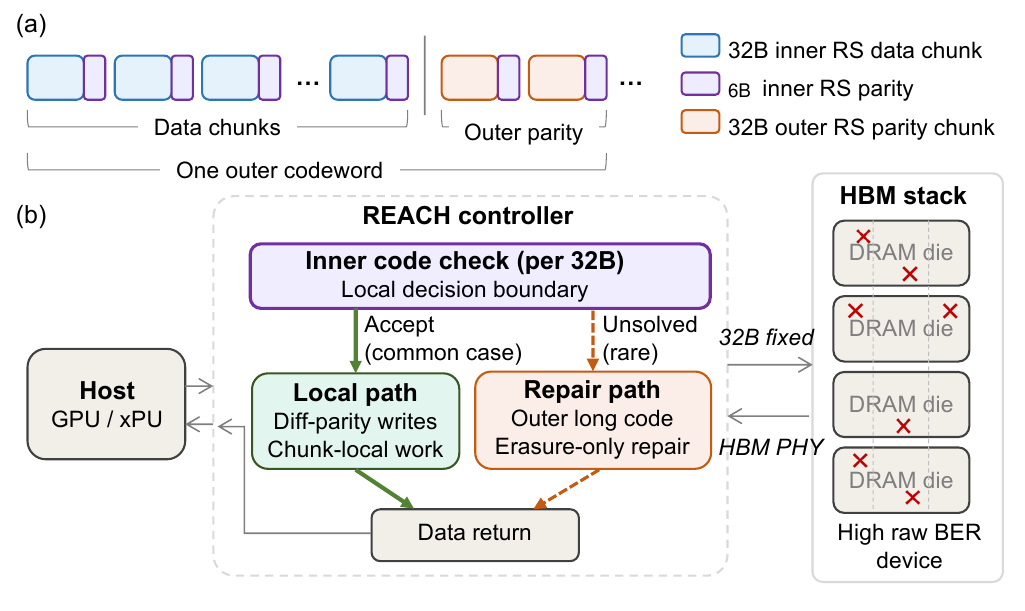}
    \caption{REACH coding and controller organization. (a) Each 32\,B data chunk carries 6\,B of inner parity, while 64 chunks and 114 outer parity symbols form a single monolithic outer codeword. (b) Accepted and corrected chunks follow the local path. Rejected chunks follow the exceptional known-erasure repair path.}
    \label{fig:reach-overview}
\end{figure}

Fig.~\ref{fig:reach-overview}(a) shows the 64 data chunks, their inner parity, and the physical outer-parity beats in 1 outer codeword. Fig.~\ref{fig:reach-overview}(b) follows a request through HBM and the inner decoder. Accepted and corrected chunks complete locally. A rejected chunk enters repair with its address-derived erasure position. The paths share the channel interface and inner decoders, but only repair stores the span identifier, repair-read bitmap, staged payloads, and erasure mask.

\subsection{Common Read Workflow}
\label{sec:q1-small}

The read path begins when the channel interface places a returned 32\,B payload and its 6\,B metadata in the bounded return queue. An inner decoder processes the 38 byte symbols. Accepted data retain the returned payload, while corrected data use the decoder output. Both enter the \emph{completion merge}, which preserves the request identifier and returns a single terminal result. A rejected target instead allocates a repair buffer. The controller records its outer-code position and inserts a 16-symbol erased placeholder at that position. It then issues 71 \emph{repair reads}, comprising the remaining 63 data chunks and 8 physical parity beats. Each repair read supplies either a resolved payload or a declared erasure for rebuilding the outer codeword.

\begin{figure}[htbp]
    \centering
    \includegraphics[width=\linewidth]{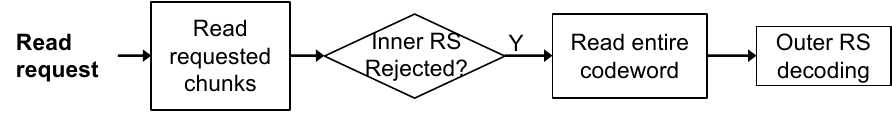}
    \caption{Common read and repair. The requested 32\,B chunk and its 6\,B metadata cross the inner decoder. Accepted and corrected targets complete immediately. A rejected target becomes a local erasure, and 63 data reads plus 8 physical parity reads complete the outer codeword for repair.}
    \label{fig:randread-flow}
\end{figure}

Fig.~\ref{fig:randread-flow} follows the target to local completion or repair. Accepted and corrected targets leave through the completion merge without outer state. Rejection retains the target as an erasure and derives 63 data and 8 parity addresses from the span base. These 71 records receive the same inner classification as the target and contribute resolved payloads or address-derived erasures. The target is not reread.

\subsection{Write and Differential-Parity Workflow}
\label{sec:q2-service}

A write checks the span's poison state and acquires an exclusive \emph{span-ordering} lock. The lock waits for older requests to the same span and holds newer reads, writes, and repairs until parity commit or poison. This ordering prevents data from one generation from being paired with parity from another. The controller resolves the touched old data through the same inner decision used by reads. If the old target is rejected, the controller assembles the outer codeword and recovers that target before continuing. The \emph{parity state} then supplies the active outer parity and its dirty, commit, and poison status to the \emph{differential-parity engine}. For sparse vectors containing only the touched chunks, RS linearity gives
\begin{equation}
\label{eq:diff-parity-sparse}
\mathbf{P}_{\mathrm{new}}=\mathbf{P}_{\mathrm{old}}\oplus
\mathrm{RS}(\mathbf{D}_{\mathrm{new}})\oplus
\mathrm{RS}(\mathbf{D}_{\mathrm{old}}).
\end{equation}
Once the old target is resolved, the differential update reads no unchanged demand data and does not rewrite unchanged chunks. A 32\,B write moves 320\,B on a parity-cache hit and 576\,B on a miss, corresponding to $10\times$ and $18\times$ amplification. The direct-outer control moves $81\times$ the payload for the same random-write case. Before device issue, the inner encoder generates 6\,B of metadata for the new target and each modified parity beat. The \emph{write-commit} block commits the coded target before the coded parity. A detected uncorrectable error (DUE) before data issue aborts the request, while a parity failure after data completion poisons the span until scrub or reinitialization.

\begin{figure}[htbp]
    \centering
    \includegraphics[width=\linewidth]{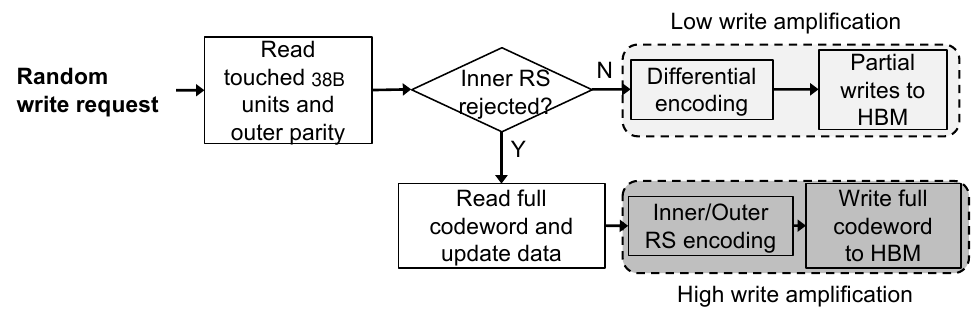}
    \caption{Differential write workflow. Exclusive span ordering covers old-data resolution, parity-state acquisition, differential encoding, metadata generation, and data-before-parity commit. A rejected old target is recovered from the outer codeword before the controller writes the new target and updated parity. Unchanged data are not rewritten.}
    \label{fig:randwrite-flow}
\end{figure}

Fig.~\ref{fig:randwrite-flow} begins when span-ordering control locks the target span. Old data and parity-fill returns cross the inner decoder. A rejected old target is first recovered from the outer codeword. A parity-cache hit supplies the 256\,B parity record, while a miss fetches and classifies its 8 beats. The differential engine combines resolved old data, new data, and parity state to generate 114 updated parity symbols. Write commit protects every changed record, completes coded data before coded parity, records poison on a terminal parity failure, and releases the lock. Complete-span recovery never implies complete-span writeback.

\subsection{Rejected-Target Repair Workflow}

The repair buffer stores the request and span identifiers, target position, repair-read completion bitmap, erasure masks, and terminal status. The \emph{outer-codeword buffer} stages the target placeholder and the 71 returned positions. Every repair read crosses the inner decoder. Accepted and corrected records contribute resolved payloads, while rejected records contribute known erasures. A full data or parity beat contributes 16 outer symbols. The final parity beat contributes only 2, and its padding does not enter the erasure budget.

\begin{figure}[htbp]
    \centering
    \includegraphics[width=\linewidth]{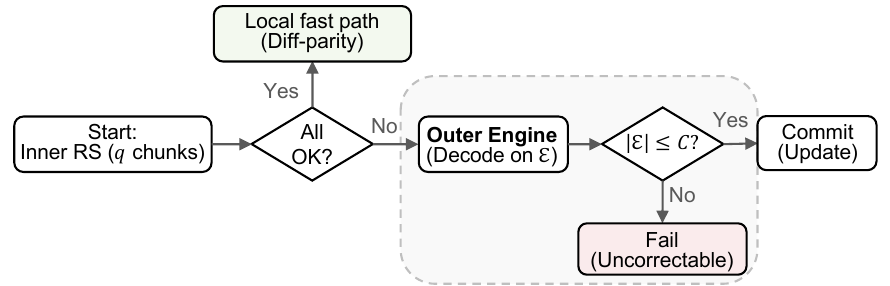}
    \caption{Known-erasure repair. A local target erasure and 71 inner-decoded repair reads assemble the complete outer codeword. At most 114 declared symbol erasures enter the outer repair engine. A successful read repair returns the reconstructed target, while a successful write repair resumes the differential update. More than 114 erasures or a failed residual check returns DUE.}
    \label{fig:erasure-decode}
\end{figure}

Fig.~\ref{fig:erasure-decode} follows the target erasure and 71 classified records into the outer buffer. A target plus 6 full rejected repair reads consumes 112 erasures, and the final 2-symbol parity tail can consume the remaining budget. A 7th full rejected repair read raises the count to 128 and returns DUE. Otherwise, the outer engine uses the assembled erasure mask to compute syndromes, magnitudes, corrections, and a residual check without a 1138-symbol Chien search. A read returns only the reconstructed 32\,B target. A write uses the recovered old target to resume its differential update.

\subsection{Controller State, Backpressure, and Completion}

The block organization assigns a unique owner to each request. \emph{Chunk-local state} holds the target payload, inner outcome, request identifier, and completion state. A repair buffer owns a rejected request while its 71 repair reads arrive. After the final read, the syndromes, erasure mask, request identity, and target position move into the post-reassembly and erasure-decoder queues, releasing the repair buffer. The outer-codeword buffer holds repair data, while the completion merge returns a local or repaired result.

\begin{figure}[htbp]
    \centering
    \includegraphics[width=\linewidth]{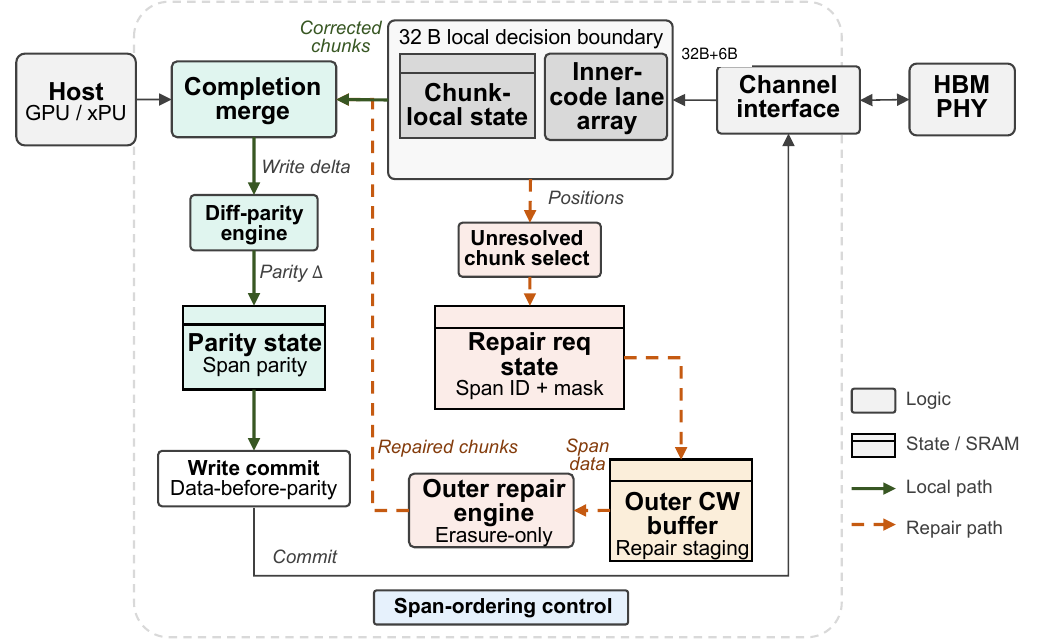}
    \caption{Block-level REACH microarchitecture. The channel interface sends every 32\,B plus 6\,B return through the inner-code lane array. Chunk-local state and the completion merge serve accepted and corrected requests. Repair-request state, the outer-codeword buffer, and the erasure-only outer engine serve rejected requests. Parity state, differential encoding, span ordering, and coded-record write commit implement data-before-parity updates. The device command scheduler and PHY remain outside the evaluated boundary.}
    \label{fig:decoder-microarch}
\end{figure}

Fig.~\ref{fig:decoder-microarch} is read from the HBM return path toward the host. The channel interface sends data and metadata to the inner lanes. Accepted and corrected targets enter chunk-local state and the completion merge. Rejected targets acquire a repair buffer, issue reads through the same interface, and place resolved payloads in the outer-codeword buffer. The span identifier and erasure mask drive outer repair, whose result rejoins the completion merge with the original request identifier. On writes, parity state feeds the differential engine, while write commit generates inner metadata and enforces data-before-parity completion under span ordering.

Finite ready-valid queues preserve request ownership and propagate backpressure. A repair buffer owns a rejected target until all 71 reads are classified, then transfers the syndrome, erasure mask, target position, and identifier to the post-reassembly and erasure-decoder queues. The buffer is released only after this transfer succeeds. Completion occurs after a local result, a repaired target with zero residual, or DUE becomes final. A parity failure after data commit records a poisoned span. Thus, no admitted request is discarded or completed from partial repair state.

\subsection{Decoupled Service Domains}

The HBM-side domain classifies returned targets and repair reads and stages completed repair contexts. A separate repair domain performs syndrome processing and erasure decoding through finite ready-valid queues. Inner decoders scale with HBM returns, while outer resources scale with rejection rate. Section~\ref{sec:evaluation} sizes both domains and tests repair bursts.

\section{Evaluation}
\label{sec:verification}
\label{sec:evaluation}
\label{sec:e2e-and-sensitivity}

The evaluation tests reliability, structured errors, LLM-decode traffic, HBM command service, and hardware cost. It combines $8\times10^6$ binary-channel samples, deterministic fault patterns, 3 public-model workloads, 10-seed Ramulator2 simulation, finite-queue stress, and 8 ASAP7-synthesized kernels. Every main result uses RS(38,32) inner codes and the monolithic 2\,KB RS(1138,1024) outer codeword. An accepted incorrect payload is silent data corruption (SDC), while a detected uncorrectable error (DUE) terminates the request. The highest HBM output BER $p_b$ is used only for controller traffic and hardware stress.

\subsection{Local Reliability Boundary}

The main model treats HBM output errors as a binary symmetric channel (BSC), where 0-to-1 and 1-to-0 flips occur independently with probability $p_b$. Each 304-bit inner codeword contains the 32\,B payload and its 6\,B metadata, so the model covers corruption in either field. The radius-2 inner decoder corrects at most 2 byte symbols and rejects every 3- and 4-symbol pattern. For traffic and hardware sizing, we conservatively treat at least 3 erroneous byte symbols as a repair:
\begin{equation}
p_{\mathrm{esc}}^{\mathrm{ub}}
=\Pr[\mathrm{Binomial}(38,1-(1-p_b)^8)\ge3].
\label{eq:pesc}
\end{equation}
The q-ary analytical estimate uses an alphabet of $q=2^8$ byte values and assumes uniformly distributed nonzero symbol differences. RS(38,32) has minimum distance 7. With radius 2, at most 4 erroneous bytes cannot cause miscorrection because 2 codewords would then be within distance 6. We therefore obtain a conservative result with no symbol-value model by treating at least 5 erroneous bytes as potentially silent. Under independent and identically distributed (IID) bit flips, $p_{\mathrm{byte}}=1-(1-p_b)^8$, yielding $\Pr[\mathrm{Binomial}(38,p_{\mathrm{byte}})\ge5]$. This result remains dependent on the IID bit-location assumption.

The BSC experiment runs the complete radius-2 decoder on the all-zero codeword, which suffices by code linearity. For each total $K$ from 5 through 12 bit flips, it samples $10^6$ patterns across 10 fixed seeds. We weight each sampled miscorrection fraction by the binomial probability of its flip count. Bonferroni-adjusted, one-sided Clopper--Pearson limits provide at least 95\% simultaneous coverage across these strata under repeated sampling. Adding the exact probability of more than 12 flips gives a conservative total-SDC confidence limit, and 10{,}000 random codewords check translation invariance.

Conditioned on target rejection, the 63 remaining data records and 7 full parity records each contribute 16 erasures when rejected. The final parity beat contains 2 logical symbols. The target, 6 full rejected repair reads, and the final tail consume all 114 available erasures. A target plus 7 full rejected repair reads exceeds the budget and causes DUE. The conservative repair probability therefore applies to the target and the 70 full repair-read groups:
\begin{equation}
P_{\mathrm{overflow},32B}^{\mathrm{ub}}
=p_{\mathrm{esc}}^{\mathrm{ub}}\,
\Pr[\mathrm{Binomial}(70,p_{\mathrm{esc}}^{\mathrm{ub}})\ge7]
\label{eq:outer-due-32b}
\end{equation}
This expression counts a rejected target followed by at least 7 rejected full repair-read groups. With correct accepted repair reads and at most 114 declared erasures, the outer decoder reconstructs the target or reports DUE after the residual check.

\begin{figure}[htbp]
  \centering
  \includegraphics[width=\linewidth]{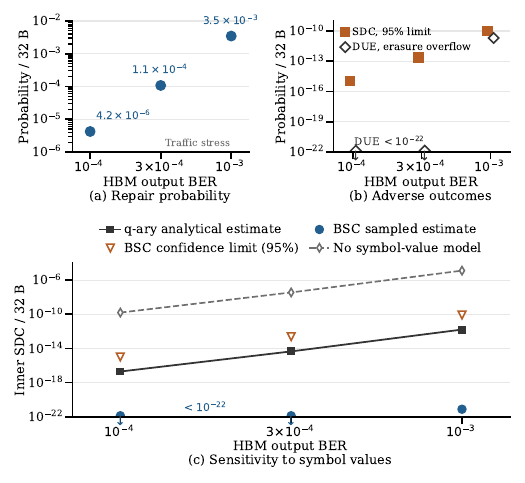}
  \caption{\textbf{Repair frequency and read outcomes per 32\,B request.} (a) Conservative repair probability. (b) Read-path SDC confidence limit and erasure-overflow DUE upper bound. The SDC limit covers the target and repair reads. Downward markers denote values below $10^{-22}$. (c) Q-ary reference, sampled BSC contribution from 5--12 bit flips, total BSC confidence limit, and the conservative result with no symbol-value model. Confidence limits are simultaneous and one-sided at 95\%.}
  \label{fig:per32b-reliability}
\end{figure}

The conservative read-path accounting has 3 components: the simultaneous one-sided 95\% confidence limit for an accepted-incorrect target, the corresponding confidence limit across 71 repair reads weighted by the repair probability, and the upper bound on more than 114 erasures. The first 2 terms form the read-path SDC confidence limit shown in Fig.~\ref{fig:per32b-reliability}(b). Correct local or repaired payloads complete successfully, while an accepted incorrect payload is SDC. Too many erasures or a nonzero residual returns DUE. A parity failure after data commit poisons the span and is outside this read-path probability.

Let $u_{\mathrm{BSC}}^{95}$ denote the simultaneous one-sided confidence limit for an accepted-incorrect inner result. A union bound gives the complete read-path error limit
\begin{equation}
P_{\mathrm{read,bad}}^{95}
\leq u_{\mathrm{BSC}}^{95}\bigl(1+71p_{\mathrm{esc}}^{\mathrm{ub}}\bigr)
+P_{\mathrm{overflow},32B}^{\mathrm{ub}}.
\label{eq:read-bad-bound}
\end{equation}
The first term covers the requested record. The factor of 71 covers accepted-incorrect repair reads when repair is invoked. The last term covers a declared-erasure count beyond the outer-code budget. Counting every accepted-incorrect repair read as adverse covers both errors detected by the residual check and errors that remain silent.

Fig.~\ref{fig:per32b-reliability} separates repair frequency, adverse read outcomes, and the sensitivity of inner SDC to erroneous byte values. All 3 subfigures use the same HBM output BER values.

Table~\ref{tab:reliability-range} separates the low-error example from the $10^{-3}$ hardware stress. For Qwen at $p_b=10^{-4}$, a union bound projects the BSC read-path SDC confidence limit to $2.14\times10^{-6}$ per token. This aggregation does not require independent requests. The result is a statistical upper limit, not an estimated field-event rate or qualification result.

At $p_b=10^{-3}$, radius-3 local decoding reduces rejection from $3.47\times10^{-3}$ to $2.40\times10^{-4}$ but raises the local q-ary SDC reference from $1.64\times10^{-12}$ to $8.82\times10^{-8}$. REACH uses radius 2 because code distance then forces every 3- and 4-symbol error to reject. The vendor-neutral 32\,B control has residual probability $2.93\times10^{-2}$ under its idealized 1-group rule.

Fig.~\ref{fig:per32b-reliability}(c) compares error-value assumptions and sampling uncertainty. The sampled BSC contribution contains 1 miscorrection across $8\times10^6$ cases and, after flip-count weighting, gives $7.79\times10^{-22}$ at $p_b=10^{-3}$. The unsampled tail has probability $1.79\times10^{-17}$ and enters the total-SDC confidence limit of $8.29\times10^{-11}$, not the sampled contribution. Assigning no distribution to erroneous byte values gives the conservative $1.30\times10^{-5}$ result. Radius 2 therefore favors explicit rejection over the lower repair rate and higher q-ary SDC reference of radius-3 local decoding.

All 48 endpoint metadata bits are used by the RS(38,32) parity. Adding a separate checksum under the same transport budget would require trading correction distance for detection rather than adding a free check. We retain the representative code and expose its error-value sensitivity in Fig.~\ref{fig:per32b-reliability}(c).

\begin{table}[htbp]
    \centering
    \caption{Role of the evaluated HBM output BER points.}
    \label{tab:reliability-range}
    \scriptsize
    \setlength{\tabcolsep}{1.5pt}
    \resizebox{\columnwidth}{!}{%
    \begin{tabular}{@{}lccccc@{}}
        \toprule
        Evaluation role & $p_b$ & \shortstack{Repair prob.\\per 32\,B} & \shortstack{Application BW\\at 3.35\,TB/s} & \shortstack{BSC read SDC\\95\% limit} & \shortstack{Overflow DUE\\upper bound} \\
        \midrule
        Low-error example & $10^{-4}$ & $4.23\times10^{-6}$ & 3.35\,TB/s & $1.04\times10^{-15}$ & $1.22\times10^{-34}$ \\
        Intermediate sensitivity & $3\times10^{-4}$ & $1.09\times10^{-4}$ & 3.32\,TB/s & $2.41\times10^{-13}$ & $2.40\times10^{-23}$ \\
        Hardware stress & $10^{-3}$ & $3.47\times10^{-3}$ & 2.69\,TB/s & $1.03\times10^{-10}$ & $2.08\times10^{-11}$ \\
        \bottomrule
    \end{tabular}%
    }
\end{table}

The BSC column combines accepted-incorrect targets and repair reads under the simultaneous 95\% confidence limit. Overflow DUE bounds only declared erasures beyond the outer repair budget, per 32\,B request. No row is a device qualification point.

\subsection{Structured and Clustered Errors}

\begin{figure}[htbp]
    \centering
    \includegraphics[width=\linewidth]{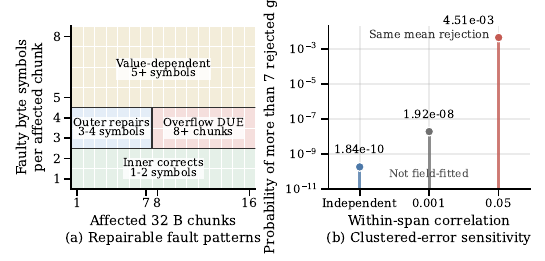}
    \caption{\textbf{Structured and clustered errors.} (a) Deterministic outcome by affected chunks and faulty byte symbols per chunk. Black lines separate the transitions after 2 and 4 faulty symbols and after 7 affected chunks. (b) Span-level probability of more than 7 rejected groups at the same mean reject count and $p_b=10^{-3}$. Correlation is a sensitivity setting, not a field-fitted parameter.}
    \label{fig:correlated-stress-rs38}
\end{figure}

Fig.~\ref{fig:correlated-stress-rs38} shows deterministic outcomes for nonuniform errors and then holds the mean reject count fixed while increasing within-span clustering.

Each coordinate in Fig.~\ref{fig:correlated-stress-rs38}(a) is a deterministic pattern rather than a probability. The horizontal coordinate counts affected 32\,B chunks in 1 outer span, while the vertical coordinate counts faulty byte symbols in each affected chunk. With at most 2 faulty symbols, each inner decoder corrects its chunk. With 3 or 4, each inner decoder rejects and exposes the chunk position to the outer code. The outer code repairs up to 7 rejected chunks and returns DUE from 8 onward. With at least 5 faulty symbols, the inner outcome depends on error values, as quantified in Fig.~\ref{fig:per32b-reliability}(c). The black lines separate these integer thresholds.

Fig.~\ref{fig:correlated-stress-rs38}(b) draws a span-specific reject probability from a Beta distribution and then draws outcomes for 64 data and 7 full-parity groups. The 2-symbol tail is separate and cannot change the full-group threshold. This Beta-Binomial construction preserves the mean count of 0.246 while increasing pairwise correlation. The span-level tail rises from $1.84\times10^{-10}$ under independence to $1.92\times10^{-8}$ at correlation 0.001 and $4.51\times10^{-3}$ at 0.05. This unconditional span sensitivity differs from the target-triggered per-request bound in Eq.~\eqref{eq:outer-due-32b}.

For mean reject probability $p$ and plotted correlation $\rho>0$, the sweep uses $\alpha=p(\rho^{-1}-1)$ and $\beta=(1-p)(\rho^{-1}-1)$. Drawing a span probability from $\mathrm{Beta}(\alpha,\beta)$ and then drawing 71 Bernoulli outcomes preserves $\mathbb{E}[E]=71p$. The point $\rho=0$ is the independent binomial limit. This construction changes only within-span concentration, so the horizontal axis does not imply a measured spatial fault process.

Each rejected data or full-parity group consumes 16 outer symbols, while the final parity beat contributes 2. These 2 tests expose the exact $C=7$ boundary without claiming a fitted row, bank, or lane fault distribution.

\subsection{Memory-Traffic-Limited Decode Rate}
\label{sec:e2e-ecc}

We use an analytical traffic model derived from public configurations for deterministic LLM decode~\cite{zhang2023llmcompass,qwen3-32b-config,glm47-flash-config,deepseek-v2-lite-config}. Each generated token reads the active bfloat16 (BF16) weights and the 8K-context key-value (KV) state, then appends 1 KV entry per layer. Qwen3-32B represents grouped-query attention (GQA). GLM-4.7-Flash and DeepSeek-V2-Lite represent compressed multi-head latent attention (MLA) with sparse experts. Table~\ref{tab:synthetic-workloads} reports the resulting traffic and controller-addressable capacity.

\begin{table}[htbp]
    \centering
    \caption{Public-configuration BF16 workload traffic at 8K context.}
    \label{tab:synthetic-workloads}
    \scriptsize
    \setlength{\tabcolsep}{1.7pt}
    \resizebox{\columnwidth}{!}{%
    \begin{tabular}{@{}llrrrrr@{}}
        \toprule
        Model & Attn. & Params total/active & All weights & Active weights & KV read/write & Addressed total \\
        \midrule
        Qwen3-32B & GQA & 32.76/31.98B & 65.50\,GB & 63.97\,GB & 2.15\,GB/262\,KB & 76.10\,GB \\
        GLM-4.7-Flash & MLA & 31.25/3.58B & 62.50\,GB & 7.16\,GB & 0.44\,GB/54\,KB & 70.81\,GB \\
        DeepSeek-V2-Lite & MLA & 15.70/2.45B & 31.40\,GB & 4.90\,GB & 0.26\,GB/31\,KB & 35.61\,GB \\
        \bottomrule
    \end{tabular}%
    }
\end{table}

The 80\,GB controller-addressable budget allocates 71.11\,GB to protected demand data and 8.89\,GB to the outer sidecar. The 6\,B metadata stored with each 32\,B endpoint record adds 15\,GB outside that address space. The resulting 95\,GB physical footprint gives a 74.85\% protected-demand storage rate. For a storage-cost comparison normalized to 1 conventional demand byte, let $f$ denote relative endpoint cost per physical byte. REACH's protected-byte storage cost is $f/0.7485$, so break-even requires a device-side reduction greater than 25.15\%. For example, a hypothetical 30\% reduction yields $0.70/0.7485=0.935$, or 6.5\% lower storage cost per protected byte. Endpoint metadata is accounted for here, separately from controller logic and power.

Weight traffic equals element size times active parameters. GQA KV traffic follows the layer count, KV-head count, head dimension, context length, and element size. MLA uses its compressed KV-latent and rotary-key dimensions. KV state is organized into 16-token PagedAttention blocks~\cite{kwon2023efficient,vllm2024blocksize}. Requests within a tensor or KV block form contiguous 32\,B runs, while physical blocks may be discontinuous. KV append bytes determine the write fraction, and consecutive appends reuse a 64-entry parity cache. Activations remain in accelerator-local storage.

\begin{figure}[htbp]
    \centering
    \includegraphics[width=\linewidth]{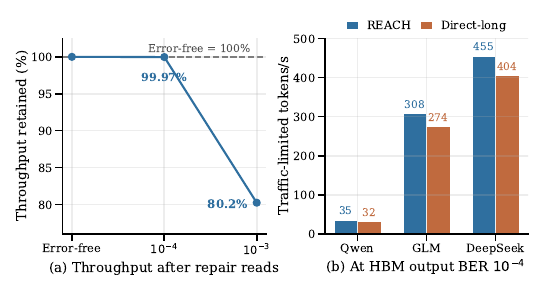}
  \caption{\textbf{Analytical LLM-decode throughput after repair traffic.} (a) Percentage of the error-free REACH token rate retained after repair reads. All 3 workloads share this curve. (b) Absolute traffic-limited rates at HBM output BER $p_b=10^{-4}$ for the 3 public-configuration BF16 workloads at 8K context. Direct-long receives ideal span coalescing. The compared designs do not have equal residual reliability.}
    \label{fig:no-extra-dq-tokens}
\end{figure}

The model uses the public H100 SXM peak HBM bandwidth of 3.35\,TB/s and an 80\,GB controller-addressable capacity budget~\cite{nvidia-h100-spec}. At $p_b=10^{-3}$, the analytical budget assigns 0.66\,TB/s to repair reads and 2.69\,TB/s to application requests, defining the hardware-provisioning point. Ramulator2 separately sustains 2.35\,TB/s total and 1.88\,TB/s of application traffic. Figs.~\ref{fig:no-extra-dq-tokens} and~\ref{fig:no-extra-dq-sensitivity} use the simulated point, while Fig.~\ref{fig:resource-capacity} and Table~\ref{tab:ppa-summary} use the analytical point. A 900\,BF16-TFLOP/s compute roof leaves all 3 models memory-bandwidth bound.

Each local decision uses $p_{\mathrm{esc}}^{\mathrm{ub}}$, and each rejected request adds 71 repair reads. The radius-3 local control uses the same 6\,B metadata without outer repair. Direct-long uses the same monolithic outer code with ideal span coalescing. The vendor-neutral 32\,B control completes native transactions without controller repair. These controls compare transfer-local protection, conventional long-code processing, and REACH's rejection-conditioned repair.

Qwen3-32B is the capacity-limiting case at 76.10\,GB after outer-sidecar allocation. The other models exercise lower active-weight traffic and compressed MLA state. Supplementary sweeps cover 2K to 32K contexts, KV blocks of 8 to 32 tokens, lower-precision weights, and 1 to 8 routed experts.

The 32\,B control partitions its hypothetical access word into 17 groups of 16 bits. Its residual probability is the binomial probability that at least 2 groups are affected. Public HBM sources do not disclose a common on-die parity matrix, so this separate control represents a capability boundary rather than a vendor implementation. The radius-3 local and 32\,B controls add no outer repair traffic. Direct-long checks every coalesced 2\,KB codeword and invokes unknown-error recovery on nonzero syndromes.

Fig.~\ref{fig:no-extra-dq-tokens} shows how repair reads affect analytical LLM-decode throughput before any policy handles terminal DUE.

Fig.~\ref{fig:no-extra-dq-tokens}(a) shows that REACH retains 99.97\% of its error-free traffic-limited rate at $p_b=10^{-4}$ and 80.2\% at the $p_b=10^{-3}$ traffic stress. The latter sizes traffic and hardware and does not define a deployment target. Under independent request outcomes, the erasure-overflow DUE upper bound projects to 4.21\% per Qwen token.

At $p_b=10^{-4}$, Fig.~\ref{fig:no-extra-dq-tokens}(b) gives REACH 35.5, 308.4, and 454.5 tokens/s for the 3 workloads, versus 31.5, 274.2, and 404.1 for direct-long. At zero error, direct-long reaches 88.9\% of REACH's traffic-limited rate because every ideally coalesced codeword carries outer parity. These values measure traffic capacity before system recovery from terminal DUE.

Ramulator2 models 5 HBM3 stacks, 80 channels, 2 pseudo-channels per channel, per-bank refresh, and HBM3-6400 timing scaled to 5234\,Mb/s per pin~\cite{luo2024ramulator2}. Balanced traffic stripes 64 data records across channels and rotates 8 parity records over the unused channels, giving every channel 8 data and 1 parity record per 10 spans. The controller uses a 1024-entry repair-read queue, a 512-entry return queue, 80 repair buffers, and 16 repair-read admissions per 2.0\,GHz cycle.

At $p_b=10^{-3}$, offered traffic is the application plus repair-read bandwidth divided by 3.35\,TB/s. The 70\% point is 2.35\,TB/s total, including 1.88\,TB/s of application requests. This offered-load percentage describes HBM command traffic and is unrelated to the 70\% controller-resource target introduced later. The selected point uses 4096 measured repairs for each of 10 seeds. The 85\% and 100\% overload diagnostics use 128 repairs for each of 2 seeds. This balanced command simulation is not workload replay or causal fault injection.

The table reports issued read-command bandwidth divided by offered traffic, the 99th-percentile time from a repair trigger to the last of its 71 reads, and the peak number of simultaneous repairs awaiting those reads. Warm-up commands crossing the fixed-interval boundary can produce 100.2\%, which equals 2.348\,TB/s and remains below 3.35\,TB/s. A point keeps up only when complete drain produces no dropped repair, admission stall, or context overflow.

\begin{table}[htbp]
    \centering
    \caption{Ramulator2 command service.}
    \label{tab:hbm-command-sim}
    \scriptsize
    \setlength{\tabcolsep}{2.5pt}
    \resizebox{\columnwidth}{!}{%
    \begin{tabular}{@{}ccccc@{}}
        \toprule
        \shortstack{Offered traffic\\(\% peak / TB/s)} & \shortstack{Issued / offered\\(\%)} & \shortstack{Repair-read latency\\(99th percentile, ns)} & \shortstack{Peak context demand\\/ 80 available} & Outcome \\
        \midrule
        70 / 2.35 & 99.5--100.2 & 324.1--347.2 & 57 & Keeps up \\
        85 / 2.85 & 86.5--88.9 & 378.0--390.2 & 70 & Falls behind \\
        100 / 3.35 & 72.7--74.5 & 537.9--620.0 & 119 & Falls behind \\
        \bottomrule
    \end{tabular}%
    }
\end{table}

At 70\% offered traffic, all 10 seeds complete, the repair-read issue queue peaks at 580 of 1024 entries, and closure holds with 10\% longer HBM timing, 128 external read-buffer entries, and age thresholds from 8 to 128 half cycles. The 85\% and 100\% points fall behind, making 70\% the highest tested traffic point that keeps up.

\subsection{Write and Span-Size Tradeoffs}
\label{sec:access-patterns}

\begin{figure}[htbp]
    \centering
    \includegraphics[width=\linewidth]{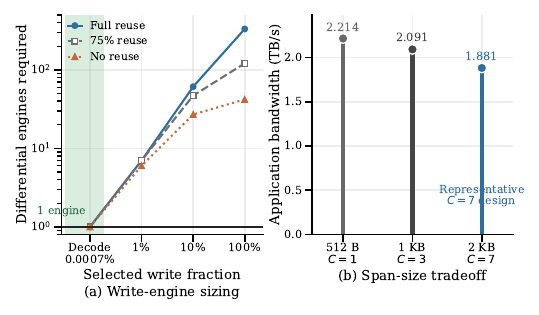}
    \caption{\textbf{Write-engine and span-size tradeoffs at 70\% HBM load.} (a) Differential-parity engines required as write fraction and parity-cache reuse change. The first category is the upper end of the evaluated LLM-decode range. (b) Application bandwidth and full-chunk repair capacity for 512\,B, 1\,KB, and 2\,KB spans at $p_b=10^{-3}$. The shorter spans are analytical counterfactuals.}
    \label{fig:no-extra-dq-sensitivity}
\end{figure}

Fig.~\ref{fig:no-extra-dq-sensitivity} sizes differential-parity engines and compares the bandwidth and repair capacity of 512\,B, 1\,KB, and 2\,KB spans. A 2.0\,GHz engine accepts 1 update every 16 cycles. At 70\% average utilization, each engine provides $87.5\times10^6$ updates/s with margin for bursts and arbitration.

The evaluated LLM-decode write fractions of 0.0004\% to 0.0007\% require 1 engine for every reuse case in Fig.~\ref{fig:no-extra-dq-sensitivity}(a). The 1\%, 10\%, and 100\% categories expose the cost of write-heavier workloads. In Fig.~\ref{fig:no-extra-dq-sensitivity}(b), 512\,B, 1\,KB, and 2\,KB spans leave 2.21, 2.09, and 1.88\,TB/s for application traffic and repair 1, 3, and 7 full chunks. The representative 2\,KB design selects the $C=7$ boundary while retaining HBM-class bandwidth.

\subsection{Full-Bandwidth Hardware Sizing}
\label{sec:ppa}

Each 32\,B application request invokes repair with probability $p_{\mathrm{esc}}^{\mathrm{ub}}$. For the read-only sizing point, application and repair traffic share the HBM bandwidth according to
\begin{equation}
\text{Application bandwidth}
=\frac{\text{HBM bandwidth}}{1+71p_{\mathrm{esc}}^{\mathrm{ub}}},
\label{eq:repair-fixed-point}
\end{equation}
At the $p_b=10^{-3}$ analytical peak-interface point, this gives 2.69\,TB/s of application bandwidth, $84.0\times10^9$ application requests/s, and $291\times10^6$ repairs/s. Workload-specific points add data, parity, and parity-fill traffic to the same budget.

We compute each hardware count from the full-bandwidth repair rate, the work performed for each repair, and the clock rate. Each post-reassembly unit requires 87 cycles. Repair-read lanes and syndrome pipelines follow the 71 repair reads. Each erasure decoder shares 114 GF$(2^{16})$ arithmetic lanes across its stages. Decoding requires 4 cycles per erasure, 1 cycle per erasure pair, and 17 fixed cycles.

For a resource that receives work at rate $\lambda_i$, performs $w_i$ cycles of work per arrival, and runs at frequency $f_i$, we select
\begin{equation}
N_i=\left\lceil\frac{\lambda_i w_i}{0.70f_i}\right\rceil.
\label{eq:resource-count}
\end{equation}
The 0.70 denominator is the stated average-utilization target. Applying this rule to $291\times10^6$ repairs/s gives 19 post-reassembly units for 87 cycles per repair, 15 repair-read lanes for 71 reads per repair, 20 syndrome pipelines for 72 positions per repair, and 30 erasure decoders for the 106.5-cycle mean service time. We use 16 repair-read lanes to align issue bandwidth with the channel organization. The inner path requires 75 lanes by rate and uses 80, matching the 80 HBM channels. These rounded counts produce the reported 80, 19, 16, 20, and 30 organization.

\begin{figure}[htbp]
    \centering
    \includegraphics[width=\linewidth]{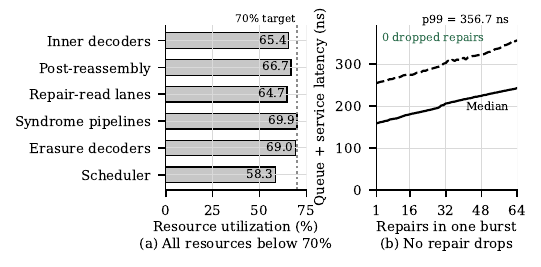}
    \caption{\textbf{Full-bandwidth repair-path capacity.} (a) Every modeled resource remains below the 70\% average-utilization target at 3.35\,TB/s and $p_b=10^{-3}$. (b) After all 71 repair reads return, the 99th-percentile queue-plus-service latency remains below 357\,ns for bursts of up to 64 repairs, with no dropped repair. Table~\ref{tab:hbm-command-sim} separately reports HBM command service and repair-read latency.}
    \label{fig:resource-capacity}
\end{figure}

The HBM-side model runs at 2.0\,GHz with 80 inner decoders, 80 repair buffers, 19 post-reassembly units, and 16 repair-read lanes. The 1.5\,GHz repair domain uses 20 syndrome pipelines and 30 erasure decoders. Finite ready-valid queues separate the domains. We target at most 70\% average utilization so that bursts, arbitration, and clock-domain crossings do not immediately create backpressure. This 70\% hardware target is a design choice and is independent of the 70\% HBM load used in the Ramulator2 simulation. Fig.~\ref{fig:resource-capacity}(a) reports utilization at the full 3.35\,TB/s interface. Fig.~\ref{fig:resource-capacity}(b) begins after all repair reads return and reports queue-plus-service latency under bursty arrivals.

Legal repairs require 106.5 repair-domain cycles on average. Fig.~\ref{fig:resource-capacity}(a) keeps every modeled block below the 70\% target at full HBM bandwidth. Fig.~\ref{fig:resource-capacity}(b) injects simultaneous groups of 1 to 64 completed repairs at the same mean rate. Every admitted repair completes, with maximum p99 post-read latency of 356.9\,ns, repair-queue depth of 64, and erasure-decoder or result-queue depth of 10. Repairs with 112 and 114 erasures require 521 and 530 cycles.

At $p_b=10^{-3}$ and full HBM bandwidth, inner decoders, post-reassembly units, repair-read lanes, syndrome pipelines, and erasure decoders reach 65.4\%, 66.7\%, 64.7\%, 69.9\%, and 69.0\% utilization. Scheduling reaches 58.3\%. The selected counts of 80, 19, 16, 20, and 30 therefore keep every modeled block within the 70\% hardware target.

We synthesize 8 parameterized SystemVerilog kernels with ASAP7 cells. Five REACH kernels implement the inner decoder, syndrome pipeline, erasure-decoder GF lane, differential-parity engine, and scheduler. Three direct-long kernels implement recursive Chien search, key-equation update, and Forney magnitude operations, with pre-layout paths of 322, 344, and 349\,ps. The area model combines mapped kernels with replica counts, explicit storage, and shared control and integration assumptions. The modeled common read path classifies a returned chunk in 13 HBM-side cycles, or 6.5\,ns at 2.0\,GHz, before local completion.

Table~\ref{tab:ppa-summary} compares both organizations at the analytical 2.69\,TB/s application point. Direct-long receives ideal 2\,KB coalescing and validates every span. Its average-error organization uses 90 syndrome pipelines, 143 key-equation paths, 1{,}065 key-equation lanes, 25{,}706 Chien-search lanes, and 713 Forney lanes. Full-radius sizing raises the last 3 counts to 10{,}117, 81{,}077, and 6{,}286.

The mean-work direct-long count weights key-equation, Chien-search, and Forney work by its conditional symbol-error distribution. This is the main comparison with REACH's mean-work provisioning. The full-radius sensitivity sets both provisioning and decoding activity to 57 unknown symbol errors, the correction radius of the 114-parity-symbol code. Both cases use mapped recursive Chien, key-equation, and Forney elements under shared ASAP7 and throughput assumptions. The comparison describes this conventional architecture, without assuming equal residual reliability or an algorithm-independent lower bound.

REACH uses 20 syndrome pipelines at the repair rate. It requires no long-code position-search lanes because inner rejection supplies the erasure positions before outer recovery begins.

The low, nominal, and high area scenarios apply the same assumptions to both organizations. SRAM periphery factors are 1.2$\times$, 1.5$\times$, and 2.0$\times$. Control allowances are 10\%, 25\%, and 50\%, followed by integration factors of 1.15$\times$, 1.30$\times$, and 1.50$\times$. Power sensitivity scales total dynamic power, including clock-pin power, with fixed leakage. The composition excludes endpoint logic, the PHY, clock-tree closure, and complete physical routing.

\begin{table}[htbp]
    \centering
    \caption{Nominal controller cost at the analytical 2.69\,TB/s application point.}
    \label{tab:ppa-summary}
    \scriptsize
    \setlength{\tabcolsep}{1.5pt}
    \resizebox{\columnwidth}{!}{%
    \begin{tabular}{@{}lcc@{}}
        \toprule
        \multicolumn{3}{c}{(a) Same-bandwidth nominal comparison} \\
        Organization & \shortstack{Area (mm$^2$)\\Ratio to REACH} & \shortstack{Modeled power (W)\\Ratio to REACH} \\
        \midrule
        \textbf{REACH} & \textbf{5.19 / 1.00$\times$} & \textbf{3.70 / 1.00$\times$} \\
        Direct-long, mean decode work & 11.73 / 2.26$\times$ & 8.73 / 2.36$\times$ \\
        Direct-long, full radius & 17.16 / 3.31$\times$ & 12.20 / 3.30$\times$ \\
        \textbf{REACH reduction vs. mean} & \textbf{55.8\%} & \textbf{57.7\%} \\
        \midrule
        \multicolumn{3}{c}{(b) REACH area breakdown} \\
        Component & Area (mm$^2$) & Share (\%) \\
        \midrule
        Inner decoders & 0.78 & 15 \\
        Syndrome pipelines & 1.29 & 25 \\
        Erasure decoders & 0.48 & 9 \\
        Write and scheduling logic & 0.051 & 1 \\
        State and storage & 0.60 & 12 \\
        Control and integration & 2.00 & 38 \\
        \textbf{Total} & \textbf{5.19} & \textbf{100} \\
        \bottomrule
    \end{tabular}%
    }
\end{table}

Across the low-to-high area assumptions, REACH spans 3.90--7.62\,mm$^2$, direct-long with mean decode work spans 9.09--16.36\,mm$^2$, and full-radius direct-long spans 13.32--23.88\,mm$^2$. Scaling total dynamic power by 0.5$\times$ to 2$\times$ gives 1.85--7.39\,W, 4.37--17.45\,W, and 6.11--24.39\,W. At nominal assumptions and mean decode work, REACH uses 55.8\% less area and 57.7\% less modeled power. The full-radius sensitivity gives reductions of 69.8\% and 69.7\%, respectively. REACH reduces hardware demand by invoking outer recovery at the repair rate and removing long-code position-search lanes from that path.

The 446\,ps slowest inner-lane pipeline stage meets the 500\,ps HBM-side block target, and every mapped repair kernel meets the 667\,ps repair-domain target. These pre-layout diagnostics establish the timing scale used in the analytical composition and are distinct from the 13-cycle common-path latency.

\subsection{Design Implications}

The results yield 3 system implications. First, the inner radius controls both reliability and hardware demand. Correcting 1 or 2 erroneous bytes keeps requests local, while rejection supplies the position needed by outer repair. A larger radius reduces repair traffic but increases the q-ary SDC reference in Fig.~\ref{fig:per32b-reliability}. The inner decision therefore determines the recovery work that the controller must provision.

Second, ordinary returns and exceptional repair require different provisioning rates. Inner lanes scale with HBM return bandwidth, while repair buffers and outer decoders scale with rejection probability and the 71-read repair cost. This separation supports the analytical full-interface target. The independent 70\% resource-utilization target reserves margin for bursts, arbitration, and clock-domain crossings.

Third, sequential weight and KV reads and sparse KV appends favor long-span protection. Fig.~\ref{fig:no-extra-dq-sensitivity} quantifies the additional differential engines needed as writes increase or parity reuse falls. Shorter spans reduce repair traffic, while the evaluated 2\,KB span provides greater erasure capacity. These tradeoffs explain the read-dominated LLM-decode specialization and guide provisioning for adjacent workloads.

\section{Related Work and Discussion}
\label{sec:related}

REACH builds on established coding primitives and our prior work~\cite{11185353}, which used per-chunk CRC detection to gate long-span recovery and differential parity for writes. REACH extends this approach with local correction and explicit rejection for both requested data and repair reads, then carries address-derived erasures through reassembly, backpressure, completion, ordered commit, poison handling, and dual-clock service. Local correction reduces the repair demand, while shared target and repair-read classification provides a consistent erasure interface. The evaluation adds binary-channel sampling, Ramulator2 command simulation, and a same-throughput hardware comparison based on synthesized kernels.

\noindent\textbf{Memory reliability.}\quad
DRAM scaling has motivated on-die, interface, rank, scrub, remapping, and retirement mechanisms~\cite{liu2013experimental,kim2014flipping,jedec-jesd79-5d,JEDEC_JESD238B01_2025,gurumurthi2021hbm3,synopsys_hbm3,nair2016xed,dell1997white,nvidia_row_remapping2023,udipi2012lot,yoon2010virtualized}. Field observations motivate our deterministic templates, although unavailable physical mapping prevents a fitted device distribution~\cite{wu2024removing}. Cerberus coordinates protection layers, while CXL-ECC, ASPA, and ReScue operate under different interfaces and service contracts~\cite{kim2026cerberus,liu2025cxl,li2026aspa,song2026rescue}.

\noindent\textbf{Controller-managed protection.}\quad
Storage and phase-change-memory controllers apply strong ECC under different latency and bandwidth contracts~\cite{zhao2013ldpc,qureshi2009enhancing}, while logic-layer memory offers greater coding flexibility~\cite{pawlowski2011hybrid}. NAND demonstrates how controller protection can preserve usable capacity as device reliability changes. REACH follows this principle while addressing HBM's finer transactions and higher bandwidth. Stronger local codes remain bounded by per-request metadata, while conventional long codes check each span and locate errors in corrupted codewords. REACH supplies unresolved chunk positions before exceptional repair.

\noindent\textbf{Customized HBM.}\quad
NVHBM places its memory controller in the HBM stack and uses a custom PHY, illustrating active co-design of controller placement and interfaces~\cite{nvidia-nvhbm-2026}. REACH evaluates an accelerator-side extension with a co-designed endpoint record, using the native in-package path and existing scheduler and PHY. NVHBM motivates customized reliability contracts without prescribing REACH's placement.

The storage-cost model in Section~\ref{sec:e2e-ecc} requires a device-side benefit exceeding the coding overhead. Actual manufacturing cost and endpoint implementation remain separate design questions.

\noindent\textbf{System contract.}\quad
REACH uses the co-designed endpoint mode in Table~\ref{tab:sideband-feasibility}. Atomic data and metadata enable the inner decision, while physical striping exposes the parallel command resources required by repair. DUE terminates the controller request. A surrounding system can add reread, scrub, retirement, or persistent-fault recovery without changing the inner-rejection and known-erasure workflow.

Alternative outer codes can preserve address-derived erasure recovery, with parity, buffering, and decoding resources resized for their service cost.


%

\section{Conclusion}
\label{sec:conclusion}
REACH targets read-dominated AI inference, where sequential weight and KV reads preserve HBM bandwidth efficiency and writes remain sparse. Within this workload regime, REACH separates ordinary 32\,B service from exceptional long-span recovery. A short inner code corrects common errors and converts unresolved chunks into address-derived erasures, so the controller invokes the monolithic outer code only after rejection. Differential parity and ordered commit handle the sparse writes without rewriting unchanged data.

Ramulator2 sustains 70\% HBM load, corresponding to 1.88\,TB/s of application bandwidth after repair reads. A separate full-interface model provisions 2.69\,TB/s of application traffic at the highest HBM output BER stress. At this target, the nominal synthesis-anchored composition uses 55.8\% less controller area and 57.7\% less modeled power than the mean-work direct-long design. These results show how regular AI-inference traffic can support strong long-span protection without placing span recovery on ordinary demand service.



\bibliographystyle{IEEEtran}
\bibliography{reference}

\begin{IEEEbiographynophoto}{Rui Xie}
received the B.E. degree from the Southern University of Science and Technology, China, and the Ph.D. degree in electrical, computer, and systems engineering from Rensselaer Polytechnic Institute, Troy, NY, USA, in 2026. His research focuses on efficient memory systems for large language models.
\end{IEEEbiographynophoto}

\vspace{-8pt}
\begin{IEEEbiographynophoto}{Yunhua Fang}
is pursuing the Ph.D. degree in electrical, computer, and systems engineering at Rensselaer Polytechnic Institute, Troy, NY, USA. He received the B.S. degree in computer science from the University of California, Davis. His research focuses on memory systems for AI infrastructure.
\end{IEEEbiographynophoto}

\vspace{-8pt}
\begin{IEEEbiographynophoto}{Asad Ul Haq}
is pursuing the Ph.D. degree in electrical, computer, and systems engineering at Rensselaer Polytechnic Institute, Troy, NY, USA. He received the B.S. degree in electrical engineering from the National University of Sciences and Technology, Pakistan. His research focuses on CXL-based computational memory systems.
\end{IEEEbiographynophoto}

\vspace{-8pt}
\begin{IEEEbiographynophoto}{Linsen Ma}
received the B.S., M.S., and Ph.D. degrees in electrical engineering from Rensselaer Polytechnic Institute, Troy, NY, USA, completing the Ph.D. in 2025. His doctoral research focused on data management over computational storage.
\end{IEEEbiographynophoto}

\vspace{-8pt}
\begin{IEEEbiographynophoto}{Sanchari Sen}
received the B.Tech. degree from IIT Kharagpur, India, and the Ph.D. degree in electrical and computer engineering from Purdue University, West Lafayette, IN, USA. She is a Staff Research Scientist at the IBM T. J. Watson Research Center, Yorktown Heights, NY, USA. Her interests include architecture and compiler design for AI accelerators and approximate computing.
\end{IEEEbiographynophoto}

\vspace{-8pt}
\begin{IEEEbiographynophoto}{Swagath Venkataramani}
received the Ph.D. degree in electrical and computer engineering from Purdue University, West Lafayette, IN, USA, in 2016. He is a Principal Research Scientist at the IBM T. J. Watson Research Center, Yorktown Heights, NY, USA. His interests include hardware and software optimizations for machine learning and approximate computing.
\end{IEEEbiographynophoto}

\vspace{-8pt}
\begin{IEEEbiographynophoto}{Liu Liu}
is an Assistant Professor in the Department of Electrical, Computer, and Systems Engineering at Rensselaer Polytechnic Institute, Troy, NY, USA. His research interests include elastic AI computing systems and architecture design. He received the B.S. degree from the University of Electronic Science and Technology of China, and the M.S. degree in electrical and computer engineering and the Ph.D. degree in computer science from the University of California, Santa Barbara.
\end{IEEEbiographynophoto}

\vspace{-8pt}
\begin{IEEEbiographynophoto}{Tong Zhang}
(Fellow, IEEE) is a Professor in the Department of Electrical, Computer, and Systems Engineering at Rensselaer Polytechnic Institute, Troy, NY, USA. His research focuses on computer system design, with an emphasis on memory-centric computing for AI and data science. He received the B.S. and M.S. degrees in electrical engineering from Xi'an Jiaotong University, China, and the Ph.D. degree in electrical and computer engineering from the University of Minnesota, Minneapolis.
\end{IEEEbiographynophoto}

\end{document}